\documentclass[a4paper,fleqn,usenatbib,times]{mnras}

\usepackage{times}

\usepackage{graphicx}    
\usepackage{amsmath}    
\usepackage{amssymb}    
\usepackage{caption}
\usepackage[dvipsnames]{xcolor}
\usepackage{tabularx}

\title[Anisotropic diffusion in SPH]{Stable anisotropic heat conduction in smoothed particle hydrodynamics}

\author[Biriukov \& Price]{
Sergei Biriukov\thanks{E-mail: sergei.biriukov@monash.edu (SB)} and
Daniel J. Price
\\
Monash Centre for Astrophysics (MoCA) and School of Physics and Astronomy, Monash University, Vic. 3800, Australia
}

\date{Accepted 2018 December 6. Received 2018 November 20; in original form 2018 August 9.}

\pubyear{2018}

\begin{document}
\label{firstpage}
\pagerange{\pageref{firstpage}--\pageref{lastpage}}
\maketitle

\begin{abstract}
  We investigate how to simulate anisotropic heat conduction in a stable manner in Smoothed Particle Hydrodynamics. We show that the requirement for stability is that entropy must increase. From this, we deduce that methods involving direct second derivatives in SPH are unstable, as found by previous authors. We show that the only stable method is to use two first derivatives with alternating differenced and symmetric SPH derivative operators, with the caveat, that one may need to apply smoothing or use an artificial conductivity term if the initial temperature jump is discontinuous. 
  Furthermore, we find that with two first derivatives the stable timestep can be 3--8 times larger even for isotropic diffusion.
\end{abstract}

\begin{keywords}
methods: numerical --- hydrodynamics --- conduction --- diffusion
\end{keywords}


\section{Introduction}
In this paper, we aim to find stable discretisations of thermal diffusion equations of the form
\begin{equation}
  \frac{{\rm d} u}{{\rm d } t} = \frac{1}{\rho}\nabla\cdot(\boldsymbol{\kappa}\nabla T),
  \label{eq_diffusiongeneral}
\end{equation}
where $u = c_v T$ is the internal energy per unit mass, $\rho$ is the density, $T$ is the temperature, and $\boldsymbol{\kappa} \equiv \kappa_{ij}$ is a heat conduction tensor (with $\kappa_{ij}\equiv \kappa \delta_{ij}$ when the conduction is isotropic). The heat conduction tensor must be a symmetric tensor ($\kappa_{ij} = \kappa_{ji}$) with positive diagonal elements, to ensure that heat flows from hot to cold and not vice versa.

Such equations arise in astrophysics when considering heat \citep{2005_Parish_mti} or cosmic ray diffusion \citep{1975_Forman_cosmray, 2016_Girichidis_cosmray} in the presence of magnetic fields, and the anisotropic diffusion of radiation \citep{2009_petkova_springel}.

Smoothed particle hydrodynamics (SPH; for reviews see \citealt{1992_monaghan_sph,2005_monagh_overall_sph}) is a Lagrangian method for fluid dynamics. The method consists of two main steps. First, discretise matter onto a set of finite particles. For each particle $a$ we define properties such as mass $m_a$, density $\rho_a$, and position $\mathbf{r}_a$. Second, discretise the set of equations using the summation interpolant \citep{1977_monagh_first, 1977_lucy_first}
\begin{equation}
  \label{eq_summationinterpolant}
  A(\mathbf{r}_a) \equiv A_a = \sum_{b}\frac{m_b}{\rho_b} A_b W(\vert \mathbf{r}_a - \mathbf{r}_b \vert, h),
\end{equation}
where $A_a$ is the interpolated property, and $W(\vert \mathbf{r}_a - \mathbf{r}_b \vert, h) \equiv W_{ab}$ is the kernel function describing the weight assigned to neighbours. As the kernel $W_{ab}$ is the only part of (\ref{eq_summationinterpolant}) that depends on $\mathbf{r}_a$, derivatives of $A_a$ with respect to the position $\mathbf{r}_a$ simply involve the gradient of the kernel function, e.g.,
\begin{equation}
  \label{eq_kernelgradient}
  \nabla_a A_a = \sum_{b}\frac{m_b}{\rho_b}\left(A_b - A_a\right)\nabla_a W_{ab}.
\end{equation}

Discretisation of the isotropic heat equation in SPH was considered in \citet{1977_lucy_first}, but most modern implementations follow the method presented by \citet{1985_brookshaw_method} and \citet{1999_cleary_monaghan} whereby
\begin{equation}
  \frac{{\rm d} u_a}{{\rm d} t} = \sum_b \frac{m_b}{\rho_a\rho_b} \overline{\kappa}_{ab} (T_b - T_a) \frac{-2F_{ab}}{|r_{ab}|}.
  \label{eq_brookshaw}
\end{equation}
In the above, $F_{ab}$ is the scalar part of the kernel gradient, and $\overline{\kappa}$ is either an arithmetic or harmonic mean of the (scalar) conductivity between the particle pair e.g. $\overline{\kappa}_{ab} = (\kappa_a + \kappa_b)/2$.

Other methods have been proposed to reduce errors in second derivatives computed with SPH by correcting the gradient operators \citep*[e.g.][]{icspmn2w}. We do not consider such methods in this paper because they do not guarantee the conservation of energy. By contrast, all of the schemes we examine conserve energy to the precision of the timestepping algorithm.

More recently, \citet{2003_espanol_revenga} generalised the Brookshaw method to the anisotropic case (i.e. when $\boldsymbol{\kappa}$ is a tensor).
However, they found that achieving accuracy to within a few percent required at least 50 neighbours in 2D, which would be equivalent to 250 neighbours in 3D. This is costly.

\citet{2009_petkova_springel} found a more significant problem: The method proposed by \cite{2003_espanol_revenga} is unstable when the diffusion is highly anisotropic. While \citet{2009_petkova_springel} proposed an `anisotropy limiter' to fix the stability problem, \citet{2017_Hopkins_anis_dif} showed that this could lead to incorrect results.


The paper is organised as follows: In Section~\ref{sec_heatcondinshp} we outline methods for isotropic and anisotropic heat conduction in SPH, and assess their stability. This shows why the Espanol \& Revenga method is unstable --- it does not guarantee positive entropy. Table~\ref{table_all_the_stability_conditions} summarises the stability conditions. We test our ideas in Section~\ref{sec_numericaltests}, and show how to remove oscillations in the solution obtained with two first derivatives when the initial conditions are discontinuous. We summarise and draw conclusions in Section~\ref{sec_discussion} and \ref{sec_conclusion}.

\section{Heat conduction in SPH}
\label{sec_heatcondinshp}
\subsection{Methods}
\label{subsec_methods}

\subsubsection{Direct second derivatives}
\label{subsubsec_directsecondderivs}

By taking second derivatives of (\ref{eq_summationinterpolant}),
an SPH discretisation of (\ref{eq_diffusiongeneral}) in the case where $\kappa$ is isotropic is given by
\begin{equation}
  \label{eq_isotropicdirectsecondderivative}
  \frac{{\rm d } u_a}{{\rm d } t} = \sum_b \frac{m_b}{\rho_a\rho_b} \overline{\kappa}_{ab} (T_b - T_a) \nabla^2_a W_{ab}.
\end{equation}

For the anisotropic case with a tensor heat conduction coefficient $\kappa_{ij}$ we can generalise this to
\begin{equation}
  \frac{{\rm d} u_a}{{\rm d} t}  = \sum_b \frac{m_b}{\rho_a\rho_b} (T_b - T_a) \left(\overline{\kappa}^{ab}_{ij} \nabla^{i}_a \nabla^{j}_a W_{ab}\right), \label{eq_directsecond}
\end{equation}
where we assume that repeated $i$ or $j$ indices imply summation.
The second derivative of the kernel can be written in terms of the dimensionless kernel function according to \citep[e.g.][]{2010_price_vector_potential}
\begin{equation}
  \nabla^{i}_a\nabla^{j}_a W_{ab} = \frac{1}{C_\nu h^{\nu+2}} \left[ \left( f^{\prime\prime} - f^{\prime} q^{-1} \right) \hat{r}_{ab}^{i} \hat{r}_{ab}^{j} + f^{\prime} q^{-1} \delta^{ij} \right].
\end{equation}
We refer to this method as the `direct second derivative' operator for heat conduction.

\subsubsection{Brookshaw method for isotropic heat conduction}
\label{subsubsec_brookshawmethod}

\cite{1985_brookshaw_method} showed that the direct second derivative operator (\ref{eq_isotropicdirectsecondderivative}) could be very inaccurate, explaining the errors reported by \cite{1977_lucy_first}. Instead, Brookshaw proposed to use the first derivative of the kernel function to compute the second derivative. 
The method is equivalent to defining a new kernel function $Y$, such that \cite[e.g.][]{2012_price_kernels}
\begin{equation}
  \nabla^2 Y_{ab} \equiv \frac{-2 F_{ab}}{|r_{ab}|},
\end{equation}
where $F_{ab}$, in terms of the dimensionless kernel function $f(q)$, is
\begin{equation}
F_{ab} = \frac{C_\nu}{h^{\nu+1}}f^\prime(q),
\end{equation}
with $q \equiv \vert r_{ab} \vert /h$, $C_\nu$ is the kernel normalisation constant, $h$ is the smoothing length and $\nu$ is the number of dimensions.
The operator for isotropic diffusion in SPH is therefore
\begin{equation}
  \frac{{\rm d} u_a}{{\rm d} t} = -\sum_b \frac{m_b}{\rho_a\rho_b} (\kappa_a + \kappa_b) (T_b - T_a) \frac{F_{ab}}{|r_{ab}|}.
  \label{eq_brookshaw}
\end{equation}

\cite{1999_cleary_monaghan} proposed an alternative form using the harmonic mean for the case where $\kappa$ is discontinuous, given by
\begin{equation}
    \frac{{\rm d} u_a}{{\rm d} t} = -\sum_b \frac{m_b}{\rho_a\rho_b} \frac{4\kappa_a\kappa_b}{(\kappa_a + \kappa_b)} (T_b - T_a) \frac{F_{ab}}{|r_{ab}|}.
\end{equation}
We do not consider the harmonic mean in this paper since \cite{2015_price_laibe} found that it could produce incorrect results in dust diffusion problems.

\subsubsection{Espanol \& Revenga: Anisotropic heat conduction}
\label{subsubsec_esanolrevengamethod}

\cite{2003_espanol_revenga} generalised the method further to be applicable to anisotropic diffusion. In this, more general case, they showed that the correct expression is
\begin{equation}
  \frac{{\rm d} u_a}{{\rm d} t} = - \sum_b \frac{m_b}{\rho_a\rho_b} T_{ba} \overline{\kappa}_{ij}^{ab} \left[(\nu + 2)\hat{r}^{i}_{ab}\hat{r}^{j}_{ab} - \delta^{ij} \right] \frac{F_{ab}}{|r_{ab}|}.
  \label{eq_esprevanisdiff}
\end{equation}
where $T_{ba} \equiv T_b - T_a$. Their method is equivalent to defining
\begin{equation}
  F_{ab}^{ij} = - \left[(\nu + 2)\hat{r}^{i}_{ab}\hat{r}^{j}_{ab} - \delta^{ij}\right] \frac{F_{ab}}{|r_{ab}|},
  \label{eq_kernelesprev}
\end{equation}
such that the anisotropic discretisation of (\ref{eq_diffusiongeneral}) is given by
\begin{equation}
  \frac{{\rm d} u_a}{{\rm d} t} = \sum_b \frac{m_b}{\rho_a\rho_b} T_{ba} \overline{\kappa}_{ij}^{ab} F_{ab}^{ij}.
  \label{eq_esprevanisdiff}
\end{equation}

\subsubsection{Petkova \& Springel: Anisotropy-limited diffusion}
\label{subsubsec_petkovaspringelmethod}

\cite{2009_petkova_springel} showed that the method proposed by \cite{2003_espanol_revenga} is unstable when the diffusion is highly anisotropic. To stabilise the method, \cite{2009_petkova_springel} proposed to modify the operator according to
\begin{equation}
  \tilde{F}_{ab}^{ij} = \gamma \kappa_{ij}^{ab}F_{ab}^{ij} + \frac{1 - \gamma}{3} \delta^{ij}.
\end{equation}
In order to obtain a stable solution in 3D, one should use at least $\gamma=\frac{2}{5}$. This converts the fully anisotropic problem to $\frac{2}{5}$ of anisotropic diffusion plus $\frac{3}{5}$ of isotropic diffusion.
However limiting the anisotropy of the diffusion in this way produces incorrect results, as discussed by \cite{2017_Hopkins_anis_dif}.

\subsubsection{Two first derivatives}
\label{subsubsec_twofirstderivs}

A simple alternative approach is to compute the diffusive flux explicitly before taking the gradient, i.e.
\begin{align}
  F^j &= \nabla^j T,  \\
  \frac{{\rm d} u}{{\rm d} t} &= \frac{1}{\rho}\nabla^i (\kappa_{ij} F^j). \label{eq_twofirstdiffusions}
\end{align}
We show in Section \ref{subsubsec_twofirststability} that this method is stable as long as we choose the derivative operators carefully. In particular, one should implement this `two first derivatives' method using a combination of differenced and symmetric derivative operators \citep{1999_cummins_conjoper,2012_tricco_price}. Considering the general case where the smoothing length on each particle is different, we discretise this using \citep[e.g.][]{2012_price_kernels}
\begin{align}
  F_a^j &= \frac{1}{\Omega_a \rho_a} \sum_b m_b (T_{b} - T_a) \nabla^j_a W_{ab} (h_a),  \label{eq_fdiff} \\
  \frac{{\rm d}u_a}{{\rm d}t} &= \sum_b m_b \left[
  \frac{\kappa^a_{ij} F_a^{i} \nabla_a^{j} W_{ab} (h_a)}{\Omega_a\rho_a^2} +
  \frac{\kappa^b_{ij} F_b^{i} \nabla_a^{j} W_{ab} (h_b)}{\Omega_b\rho_b^2} \right].
  \label{eq_fdiff2}
\end{align}

In the above, $\Omega$ is the term used in the evaluation of the smoothing length \cite{2002_monaghan_compressturb, 2002_springel_cosmosph}
\begin{equation}
  \Omega_a \equiv 1 + \frac{3 h_a}{\rho_a}\sum_b m_b \frac{ \partial W_{ab}(h_a) }{ \partial h_a}.
\end{equation}

For completeness, in Section~\ref{subsec_diffvssymm} we tested the derivative operators in the reverse order --- differenced after symmetric. Although we find no significant difference in the results (see Figures~\ref{fig_hc12_k100_symdif_vs_difsym_discont}~and~\ref{fig_hc12_k010_symdif_vs_difsym_discont}), the order above is necessary to conserve total energy, $E$, since
\begin{equation}
\frac{{\rm d} E}{{\rm d} t} =  \frac{{\rm d} }{{\rm d} t}  \sum_a m_a u_a = \sum_a m_a  \frac{{\rm d} u_a}{{\rm d} t}  = 0.
\end{equation}
The last sum is zero when inserting (\ref{eq_fdiff2}) because of the antisymmetry of the kernel gradient producing a double summation which is antisymmetric in the particle index $a$ and $b$ and therefore equal to zero. The above version is also efficient because it does not require adding a third loop over the particles between the density and force evaluations --- since evaluating the flux does not require prior knowledge of the density.

\begin{figure*}
  \begin{center}
    \includegraphics[width=1\linewidth]{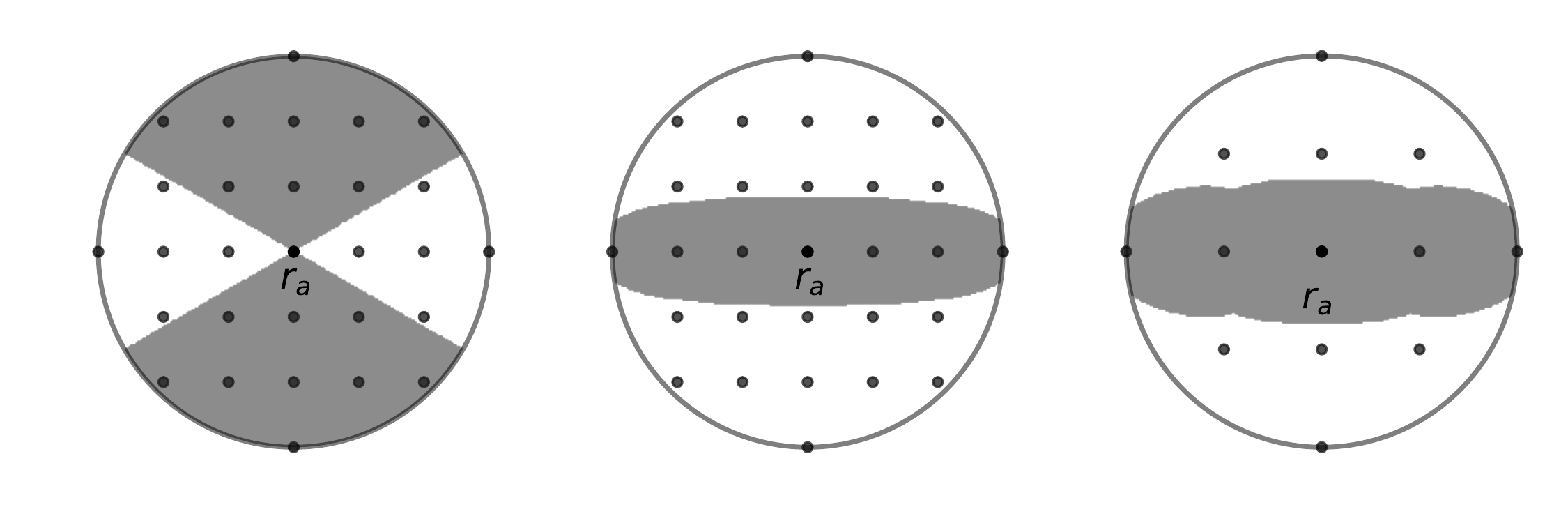}
    \hspace*{\fill}
    \begin{minipage}{.32\linewidth}
    \centering
    a) Espanol-Revenga M$_6$
    \end{minipage}
    \begin{minipage}{.32\linewidth}
    \centering
    b) Direct 2$^{nd}$ Derivative M$_6$
    \end{minipage}
    \begin{minipage}{.32\linewidth}
    \centering
    c) Direct 2$^{nd}$ Derivative M$_4$
    \end{minipage}
    \captionof{figure}{\label{fig_stabilitycircles} Neighbours within the kernel radius that contribute to anti-diffusive behaviour when simulating anisotropic diffusion. We consider uniformly distributed particles inside a compact domain for $a,b)$ quintic spline kernel and $c)$ cubic spline kernel, both with the smoothing length equal to the particle spacing. Instability regions are shown in grey for anisotropic diffusion assuming $\kappa_{xx}=1$.
    The first plot $a)$ shows this region for the Espanol-Revenga method, while plots $b,c)$ show it for the direct second derivative method.
  Grey shading indicates the region from where the neighbours add negative diffusion to a particle $a$.
   With the Espanol-Revenga method (left), the anti-diffusive region is larger, which is why this method becomes unstable faster.
   The region becomes smaller relative to the kernel support area with the quintic kernel for the direct 2$^{nd}$ derivative method, but better kernels cannot prevent instability altogether.
   }
  \end{center}
\end{figure*}

\subsection{Stability}
\label{subsec_stability}

As stated by \cite{2005_monagh_overall_sph}, the main advantage of the Brookshaw method over a direct second derivative of the SPH kernel is the guarantee of increasing entropy. For heat conduction problems, this means that heat always flows from hot to cold particles (which guarantees stability but says nothing about accuracy). To show that this is true, we need to prove that the rate of change of the total entropy $S$ is positive. For the Espanol-Revenga operator, we have
\begin{equation}
  \begin{aligned}
    &\frac{{\rm d}S}{{\rm d}t} = \sum_a m_a\frac{{\rm d}s_a}{{\rm d}t} = \\
    &\sum_a \sum_b \frac{m_a m_b}{\rho_a \rho_b}\left(\frac{1}{T_a} - \frac{1}{T_b}\right)(T_b - T_a) \overline{\kappa}_{ij}^{ab} F_{ab}^{ij}.
    \label{eq_totalentropy}
  \end{aligned}
\end{equation}

As the terms involving mass, density and temperature are all positive, positive entropy requires
$\overline{\kappa}_{ij}^{ab} F_{ab}^{ij}$
to be positive definite.

\subsubsection{Brookshaw: Isotropic heat conduction}
\label{subsubsec_brookshawstability}

When heat conduction is isotropic ($\boldsymbol{\kappa} = \kappa \delta^{ij}$), the \cite{2003_espanol_revenga} operator reduces to the \cite{1985_brookshaw_method} operator. The stability condition becomes
\begin{equation}
  \overline{\kappa}_{ii}^{ab} F_{ab}^{ii} = \kappa \frac{-2 F_{ab}}{|r_{ab}|} \geq 0.
  \label{eq_brookshawstable}
\end{equation}
Because $\kappa$ is always positive and $F_{ab}$ is negative (assuming SPH kernels where the weight decreases monotonically with radius), the kernel part (\ref{eq_brookshawstable}) and the total entropy (\ref{eq_totalentropy}) are always positive, and therefore the method is stable.
\label{label_instabilityexplanition}

\subsubsection{Espanol \& Revenga: Anisotropic heat conduction}
\label{subsubsec_esanolrevengastability}

As the following idea is the same for more than one dimension, we will discuss the simplest case of 2D anisotropic heat conduction.
Knowing that, from the energy argument, the heat conduction tensor is a real symmetric tensor, it is always possible to find a principal axis and to represent the heat conduction tensor as a diagonal (orthotropic) tensor using spectral decomposition with respect to the principal axis.
This means that for any anisotropic heat conduction problem we can find a new local coordinate axis in which $\boldsymbol{\kappa}$ is a diagonal tensor.
Thus, to have positive total entropy (\ref{eq_totalentropy}) in 2D, we need to satisfy the inequality
\begin{equation}
  -\kappa_{xx}[(\nu+2)\hat{r}^x\hat{r}^x - 1)] -\kappa_{yy}[(\nu+2)\hat{r}^y\hat{r}^y - 1] \geq 0.
  \label{eq_esprevinstabilitycondition}
\end{equation}
To satisfy the inequality (\ref{eq_esprevinstabilitycondition}), the ratio of components should be
\begin{equation}
  \frac{\kappa_{xx}}{\kappa_{yy}}\in\left[\frac{1}{\nu+1}; \nu+1\right].
\end{equation}
As in practice this ratio can be arbitrary, anti-diffusive behaviour is possible. To illustrate this, the grey shading in Figure \ref{fig_stabilitycircles} shows the regions inside the kernel compact support radius from which neighbouring particles contribute anti-diffusive terms. The instability grows from these regions resulting in negative temperatures.

\subsubsection{Direct second derivatives}
\label{subsubsec_directsecondderivsstability}

While we found this approach to be more stable than Espanol-Revenga method both practically and theoretically, after the same analysis as in Section~\ref{label_instabilityexplanition} we arrive at a similar condition for stability as for the Espanol-Revenga operator, namely
\begin{equation}
  \begin{aligned}
    &-\kappa^{xx}[(f^{\prime\prime} - f^{\prime} q^{-1})
    \hat{r}^{x}\hat{r}^{x} + f^{\prime} q^{-1}] \\
    &-\kappa^{yy}[(f^{\prime\prime} - f^{\prime} q^{-1})
    \hat{r}^{y}\hat{r}^{y} + f^{\prime} q^{-1}] \geq 0,\\
  \end{aligned}
  \label{eq_dirderinstabilitycondition}
\end{equation}
or in the isotropic case
\begin{equation}
  \begin{aligned}
    &-\kappa [(f^{\prime\prime} - f^{\prime} q^{-1}) + \nu f^{\prime} q^{-1}] \geq 0.
  \end{aligned}
  \label{eq_dirderinstabilityconditionisotropic}
\end{equation}

This means that for direct second derivatives with standard kernels there is always a region in which anti-diffusive behaviour occurs. These regions are shown for different kernels in grey in Figure \ref{fig_stabilitycircles} for the test case described in Section \ref{subsec_difftest3dconstkappa} where $\kappa^{xx} = 1$ and $\kappa^{yy}=0$.

\subsubsection{Two first derivatives}
\label{subsubsec_twofirststability}

\begin{table}
  \begin{tabular}{ | l | c | c | }
    \hline
    Method & Isotropic stability & Anisotropic stability  \\
    \hline
    Brookshaw/Espanol-Revenga & Always & If Eq.~\ref{eq_esprevinstabilitycondition} satisfied \\
    Direct second derivatives
    & If Eq.~\ref{eq_dirderinstabilityconditionisotropic} satisfied &
If Eq.~\ref{eq_dirderinstabilitycondition} satisfied \\
    Two first derivatives & Always & Always \\
    \hline
  \end{tabular}
  \caption{
  \label{table_all_the_stability_conditions}
  Summary of stability conditions for anisotropic diffusion in SPH using different methods.}
\end{table}

We compute the entropy evolution using
\begin{equation}
  \begin{aligned}
    \frac{{\rm d}S}{{\rm d}t} &= \sum_a m_a \frac{{\rm d} s_a}{{\rm d}t} = \sum_a m_a \frac{1}{T_a} \frac{{\rm d} u_a}{{\rm d}t} =\\
    & \sum_a m_a \frac{1}{T_a} \sum_b m_b \left[
    \frac{\kappa^a_{ij} F_a^{i} \nabla_a^{j} W_{ab}(h_a)}{\Omega_a\rho_a^2} +
    \frac{\kappa^b_{ij} F_b^{i} \nabla_a^{j} W_{ab}(h_b)}{\Omega_b\rho_b^2} \right].
  \end{aligned}
\end{equation}
Taking into account that $\nabla_a W_{ab} = - \nabla_b W_{ba}$ we can rearrange the double summation to give
\begin{equation}
  \begin{aligned}
    \frac{{\rm d}S}{{\rm d}t} &= \sum_a \frac{m_a}{T_a T_b}\frac{\kappa^a_{ij} F_a^{i}}{ \Omega_a\rho_a^2} \sum_b m_b (T_b - T_a)\nabla_a^{j} W_{ab} (h_a).
  \end{aligned}
\end{equation}
Hence, as long as the flux is calculated using (\ref{eq_fdiff}), entropy increase is guaranteed. Importantly, stability is guaranteed \emph{independent of the choice of SPH kernel}.
This result is similar to the proof given in \cite{2017_Price_Phantom} for physical viscosity or in \cite{2012_price_kernels} for magnetohydrodynamics.

Table~\ref{table_all_the_stability_conditions} summarises the stability conditions for each of the methods discussed above.

\subsection{Timestep constraints}
For diffusion problems, the timestep requirement is given by (e.g. \citealt{1999_cleary_monaghan})
\begin{equation}
  \Delta t \leq \frac{C c_v \rho h^2}{\kappa_{ij}},
\end{equation}
where $C$ is a constant. The above is a local constraint. Since we used the same timestep for all particles, we take the minimum over all particles.

Table~\ref{table_all_the_integration_constants} gives the values of $C$ we found to be stable in our 2D tests. We obtained these values empirically, although they could, in principle, be determined analytically by a stability analysis. The constants for Brookshaw and direct second derivative methods apply only to isotropic diffusion, as the methods are unstable for anisotropic diffusion regardless of the choice of timestep, which we confirmed down to $C=0.01$.

The change of the stability constant between kernels (left to right in Table~\ref{table_all_the_integration_constants}) is not surprising. This simply follows the standard deviation of the kernel, as discussed by \cite{2012_dehnen_aly_analysis}. Even so this effect helps to mitigate the computational cost of smoother spline kernels.

The surprising aspect is the factor of 3--8 increase in timestep possible with the two first derivatives method (bottom row of Table~\ref{table_all_the_integration_constants}). We confirmed that this also holds in 3D. We attribute this to the extra stability provided by the double convolution of the kernel gradient inherent in the two first derivatives method. This gives us compelling reason to use this method even for isotropic diffusion.

\begin{table}
  \begin{tabular}{| l | c | c | c |}
    \hline
     & $M_4$ & $M_6$ & $M_8$ \\
    \hline
    Brookshaw/Espanol-Revenga & 0.15 & 0.18 & 0.2 \\
    Direct second derivatives & 0.2 & 0.35 & 0.4 \\
    Two first derivatives & 0.6 & 1.2 & 1.6 \\
    \hline
  \end{tabular}
  \caption{
  \label{table_all_the_integration_constants}
  The integration constants used in the numerical tests.}
\end{table}

\section{Numerical tests}
\label{sec_numericaltests}

\subsection{Diffusion in a slab with constant heat conduction tensor}
\label{subsec_1ddifftest}
 To test our ideas in practice, we considered diffusion in a 3D slab: $x\in[-1,1]$ with Dirichlet boundary condition, and  $y,z\in[-4 \Delta r, 4 \Delta r]$, where $\Delta r$ is the particle spacing, with periodic boundary conditions. We set up the problem using $64 \times 8 \times 8$ particles in 3D. Initially, $T_L = 1$ for $x < 0$, and $T_R = 2$ for $x > 0$. In the isotropic case the exact solution is one dimensional
 and can be approximated for some time (until diffusion hits the boundary) by the exact solution for one dimensional heat conduction in an infinite slab
\begin{equation}
  T(x,y,z,t) = T(x,t) = \frac{T_R + T_L}{2} + \frac{T_R - T_L}{2}{\rm Erf}\left(\frac{x}{\sqrt{4\kappa_{xx} t}}\right).
\end{equation}
This solution holds for anisotropic diffusion when $\kappa^{xx}$ is the only non-zero component
\begin{equation}
  \frac{\partial T}{\partial t} = \kappa_{xx} \frac{\partial^2 T}{\partial x^2}.
\end{equation}

\subsubsection{Diffusion in the direction of the heat gradient}
\label{subsec_1ddifftest1}
\begin{figure}
  \begin{center}
    \includegraphics[width=1\linewidth]{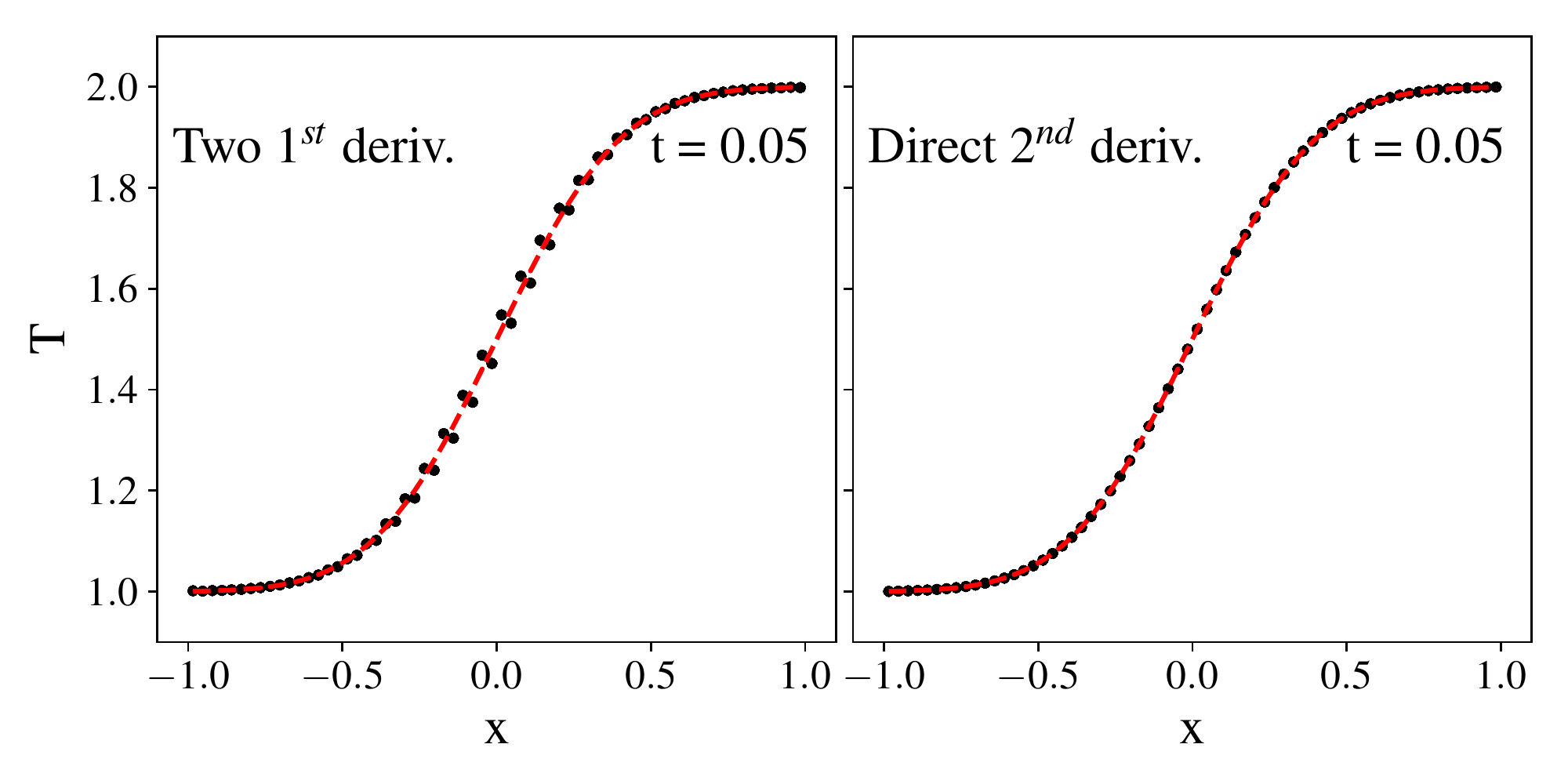}
    \captionof{figure}{Diffusion along the x-direction in a 3D slab. Black dots show the SPH particles, red dashed line shows the exact solution.  As expected, the diffusion for anisotropic case with only one non-zero component $\kappa^{xx}=1$ behaves in the same way as for isotropic diffusion with $\kappa=1$.}
    \label{fig_hc12k100}
  \end{center}
\end{figure}

Figure \ref{fig_hc12k100} shows the results of this test at a resolution of 64 particles along the x-direction, placed on a uniform lattice. We compare the two first derivatives method (left) to direct second derivatives (right), finding that both two first derivatives and direct second derivative methods produce stable results.
Using noisier kernels (M$_4$ instead of M$_6$) or higher resolution leads to numerical instability with the direct second derivative.

At this resolution and kernel (M$_6$), a `carbuncle mode' appears in the solution obtained with the two first derivatives method caused by the initially discontinuous temperature profile, while the solution obtained with direct 2$^{nd}$ derivative is less noisier. This error decreases with resolution, and can be eliminated by smoothing the initial conditions (see Section~\ref{subsec_elimcarbmode}).

\subsubsection{Diffusion perpendicular to the heat gradient}

When the only non-zero component of heat conduction tensor is $\kappa^{yy}$, but heat conduction $T(x,t)$ depends only on the $x$ component, there should be no diffusion at all.

\begin{figure}
  \begin{center}
    \includegraphics[width=1\linewidth]{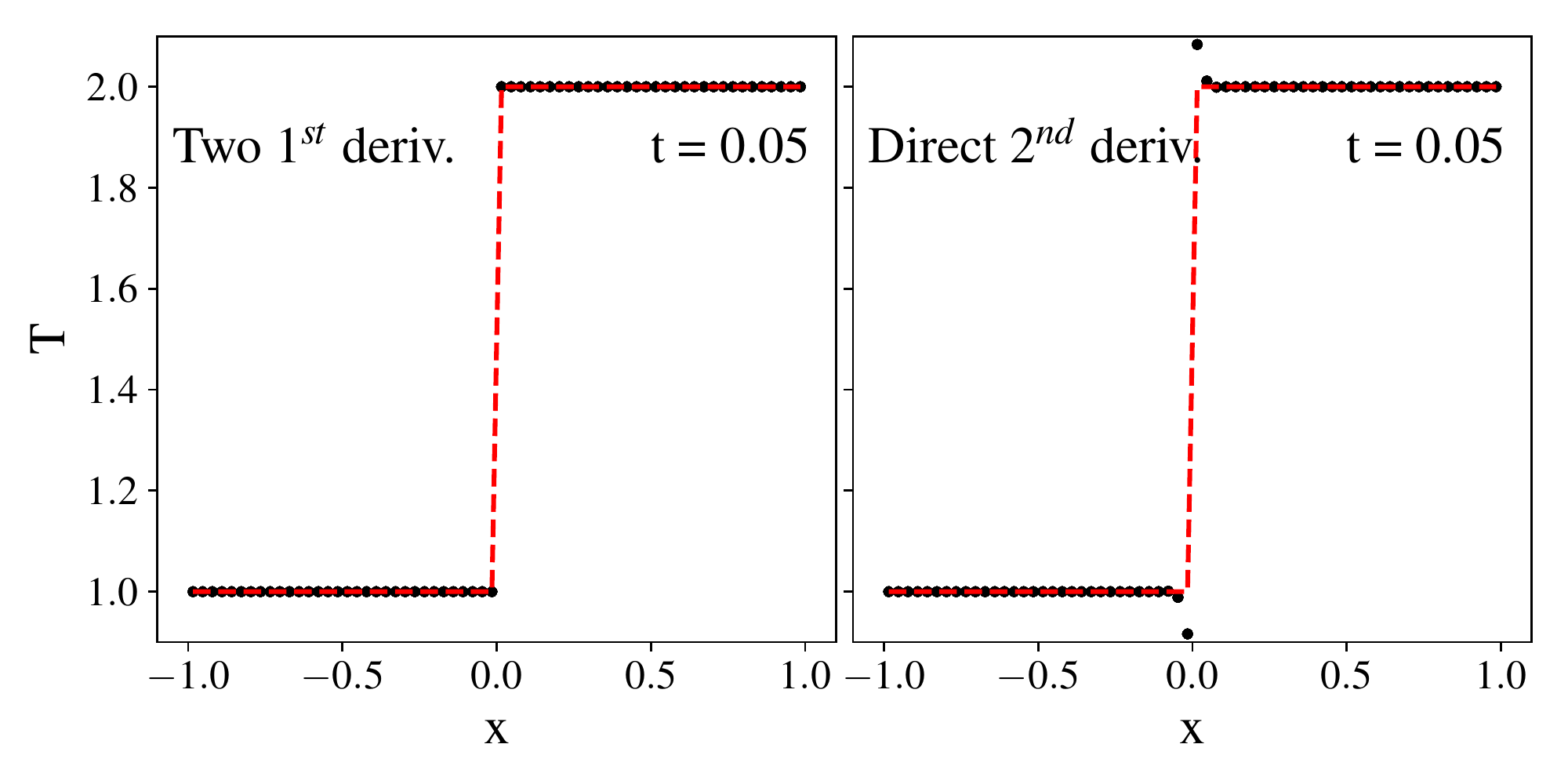}
    \captionof{figure}{\label{fig_hc12k010} Anisotropic diffusion in a 3D slab (Section~\ref{subsec_1ddifftest}). Black dots show the SPH particles, red dashed line shows the exact solution. Here, we assume $\kappa^{yy} = 1$, meaning there should be no heat conduction at all in the x-direction. The direct second derivatives method is unstable in this case, showing anti-diffusive behaviour (right panel), while the two first derivatives method remains stable and accurate (left panel).}
  \end{center}
\end{figure}

 Figure \ref{fig_hc12k010} shows the results of the diffusion test with conduction allowed only in the direction perpendicular to the heat gradient. In contrast to the previous case, two first derivatives give the correct solution (left), while the direct second derivatives method is unstable (right).

\subsubsection{Does the order of derivative operators matter?}
\label{subsec_diffvssymm}
 Comparing left and right panels in Figures~\ref{fig_hc12_k100_symdif_vs_difsym_discont}~and~\ref{fig_hc12_k010_symdif_vs_difsym_discont} shows that the order of operators (Symmetric after Differential in the left column or Differential after Symmetric in the right columns) does not affect the results. However, total energy is only exactly conserved when the symmetric operator is used in the thermal energy equation.

\subsubsection{Eliminating the carbuncle mode}
\label{subsec_elimcarbmode}

\begin{figure}
  \begin{center}
    \includegraphics[width=1\linewidth]{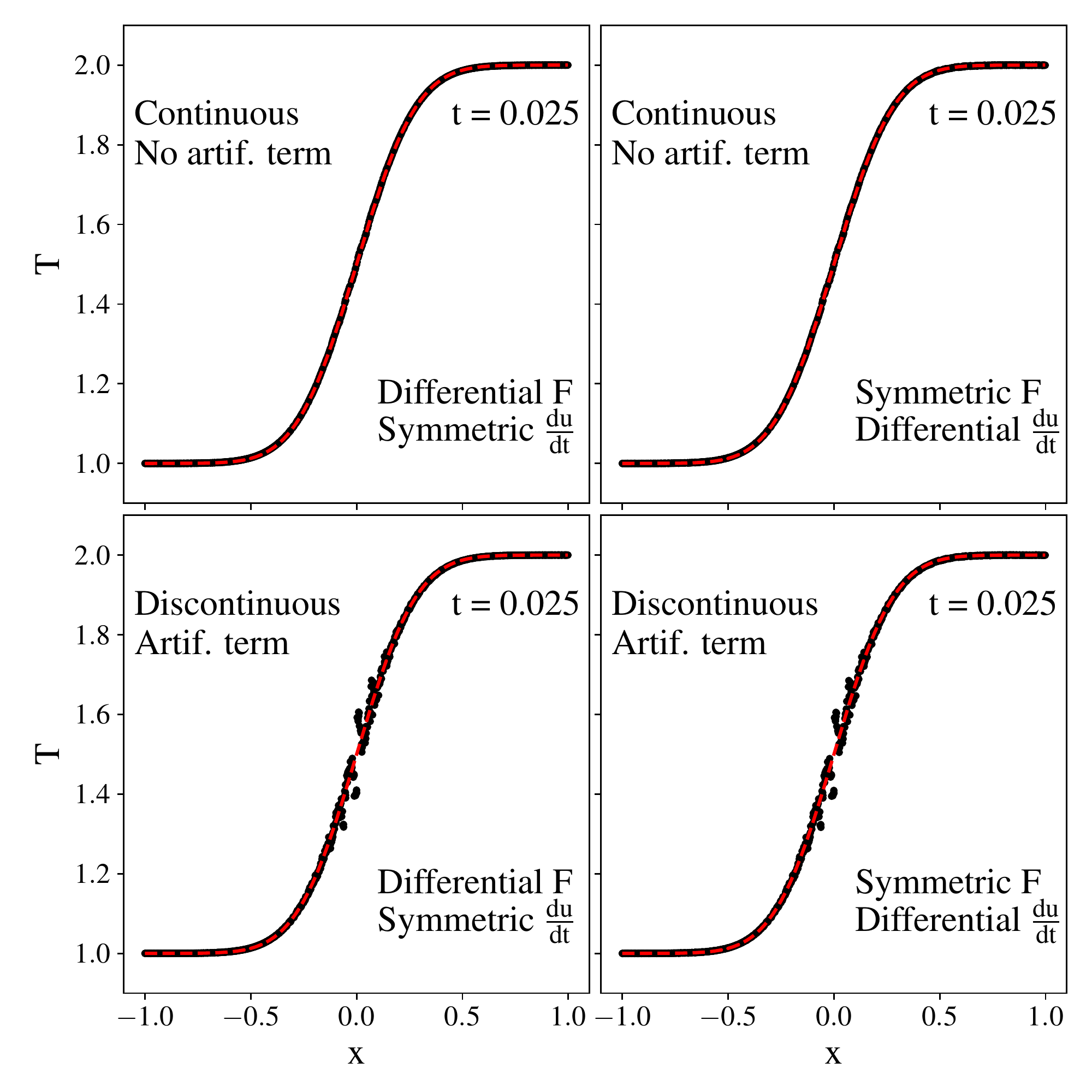}
    \captionof{figure}{
    As in Figure~\ref{fig_hc12k100} but with a glass-like initial setup.
    First, a `carbuncle mode' originates purely because of a discontinuous initial conditions, as the mode is eliminated with the continuous initial conditions (top row of this Figure) or the artificial thermal conduction term (bottom row).
    Second, the order of operators (Symmetric after Differential in the left column or Differential after Symmetric in the right columns) does not affect the final result of the diffusion operator when it is done on the same set of relaxed particles.
    }
    \label{fig_hc12_k100_symdif_vs_difsym_discont}
  \end{center}
\end{figure}

\begin{figure}
  \begin{center}
    \includegraphics[width=1\linewidth]{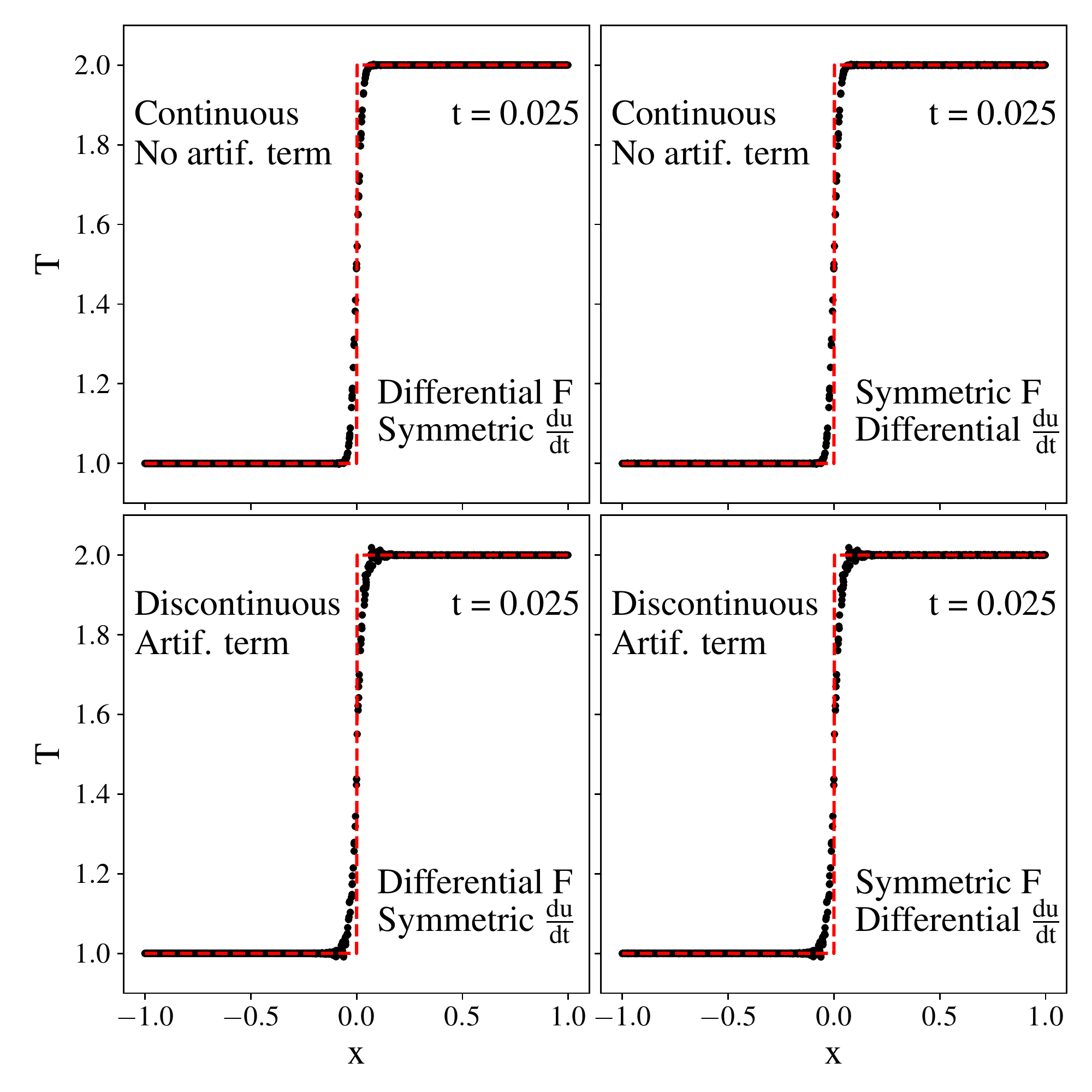}
    \captionof{figure}{
    As in Figure~\ref{fig_hc12_k100_symdif_vs_difsym_discont} but with only $\kappa_{yy}=1$.
    Although, both methods cause the solution to loose the sharp interface, that can be fixed with higher resolution (see Section~\ref{sec_convergence}).
    The order of operators still does not affect the final result.
    }
    \label{fig_hc12_k010_symdif_vs_difsym_discont}
  \end{center}
\end{figure}

The reason for the oscillations in Section~\ref{subsec_1ddifftest1} is the discontinuity in the initial conditions. One can adopt smooth initial conditions by setting the initial temperature according to
\begin{equation}
T_0  = 1.5 + 0.5 {\rm Erf}(x/L),
\end{equation}
where $L$ is the length scale over which the initial discontinuity is smoothed, which we set to twice the initial particle spacing. The top row of Figure \ref{fig_hc12_k100_symdif_vs_difsym_discont} demonstrates that the oscillations indeed vanish in this case (compare top to bottom row).

\cite{2008_Price_discontSPH} showed that, in general, one requires artificial dissipation terms whenever a discontinuity occurs in SPH.
Hence an alternative solution is to add an artificial conductivity term at the discontinuity. For shock capturing the usual term is of the form
\begin{equation}
 \left(\frac{{\rm d}u}{{\rm d}t}\right)_{\rm AC} = \sum_b m_b \alpha_{\rm u} v_{\rm sig}^{\rm u} (u_a - u_b) \frac{1}{2}\left[\frac{F_{ab}(h_a)}{\Omega_a\rho_a} + \frac{F_{ab}(h_b)}{\Omega_b\rho_b}\right],
  \label{eq_diffusion_art_term}
\end{equation}
where $\alpha_u \in [0,1]$ and $v_{\rm sig}^{u}$ is the signal speed. For hydrodynamics, \citet{2008_Price_discontSPH} proposed
\begin{equation}
v_{\rm sig}^{\rm u} = \sqrt{\frac{\vert P_a - P_b \vert}{\overline{\rho}_{ab}}},
\end{equation}
which is designed to smooth artificial pressure blips which can occur at contact discontinuities. For the purpose of this paper --- considering the heat equation in isolation --- we adopt a simpler signal speed of the form
\begin{equation}
v_{\rm sig}^{\rm u} = \sqrt{\vert u_a - u_b \vert},
\end{equation}
which could be easily generalised for diffusion of any quantity. Importantly, the addition of this term does not affect the convergence of the method, since the artificial diffusion with $v^u_{\rm sig}$ set as above is $\mathcal{O}(h^2)$. Given this it is safe to simply adopt $\alpha_u = 1$.

Figures~\ref{fig_hc12_k100_symdif_vs_difsym_discont}~and~\ref{fig_hc12_k010_symdif_vs_difsym_discont} demonstrate that both using the artificial conduction term (\ref{eq_diffusion_art_term}) or starting with the smoothed initial conditions effectively eliminate the `carbuncle mode' caused by the discontinuity.  For this figure we also placed the particles on a more realistic particle distribution, by initially placing particles randomly in the domain and relaxing them using the usual SPH equations with a damping term. We refer to this as a `glass-like' particle arrangement. 

\subsection{3D diffusion with constant heat conduction tensor}
\label{subsec_difftest3dconstkappa}

For our next test, we consider a domain $x,y,z\in[-1,1]$ with Dirichlet boundary conditions.
Assuming an initial heat distribution in the form of a delta function located at the origin and using Green functions, we can find that the solution for the isotropic equation in 3D is given by
\begin{equation}
  T(r,t) = \frac{2\pi^{-3/2}}{(\epsilon^2 + 2\kappa t)^{3/2}}
  \exp{\left(-\frac{1}{2}\frac{r^2}{\epsilon^2 + 2\kappa t}\right)}.
\end{equation}
As it is impossible numerically to start with this solution, we set the temperature at $t=0$ to be zero everywhere except a sphere of radius $\epsilon = 0.1$ around the origin where it is
\begin{equation}
  T(r,0) = \frac{2\pi^{-3/2}}{(\epsilon^2)^{3/2}}
  \exp{\left(-\frac{r^2}{2 \epsilon^2}\right)}.
\end{equation}
We also considered an anisotropic case with diffusion acting only along the $x$ axis with $\kappa_{xx}=1$. As the solution is then still the product of Green functions acting in each direction, the time evolution occurs only in the function corresponding to the nonzero component of the heat conduction tensor. The exact solution is then
\begin{equation}
  T(r,t) = \frac{2\pi^{-3/2}}{\epsilon^2(\epsilon^2 + 2\kappa t)^{1/2}}
  \exp{\left(-\frac{1}{2}
  \left[\frac{x^2}{\epsilon^2 + 2\kappa t} + \frac{y^2+z^2}{\epsilon^2}\right]\right)}.
\end{equation}

\begin{figure}
  \begin{center}
    \includegraphics[width=1\linewidth]{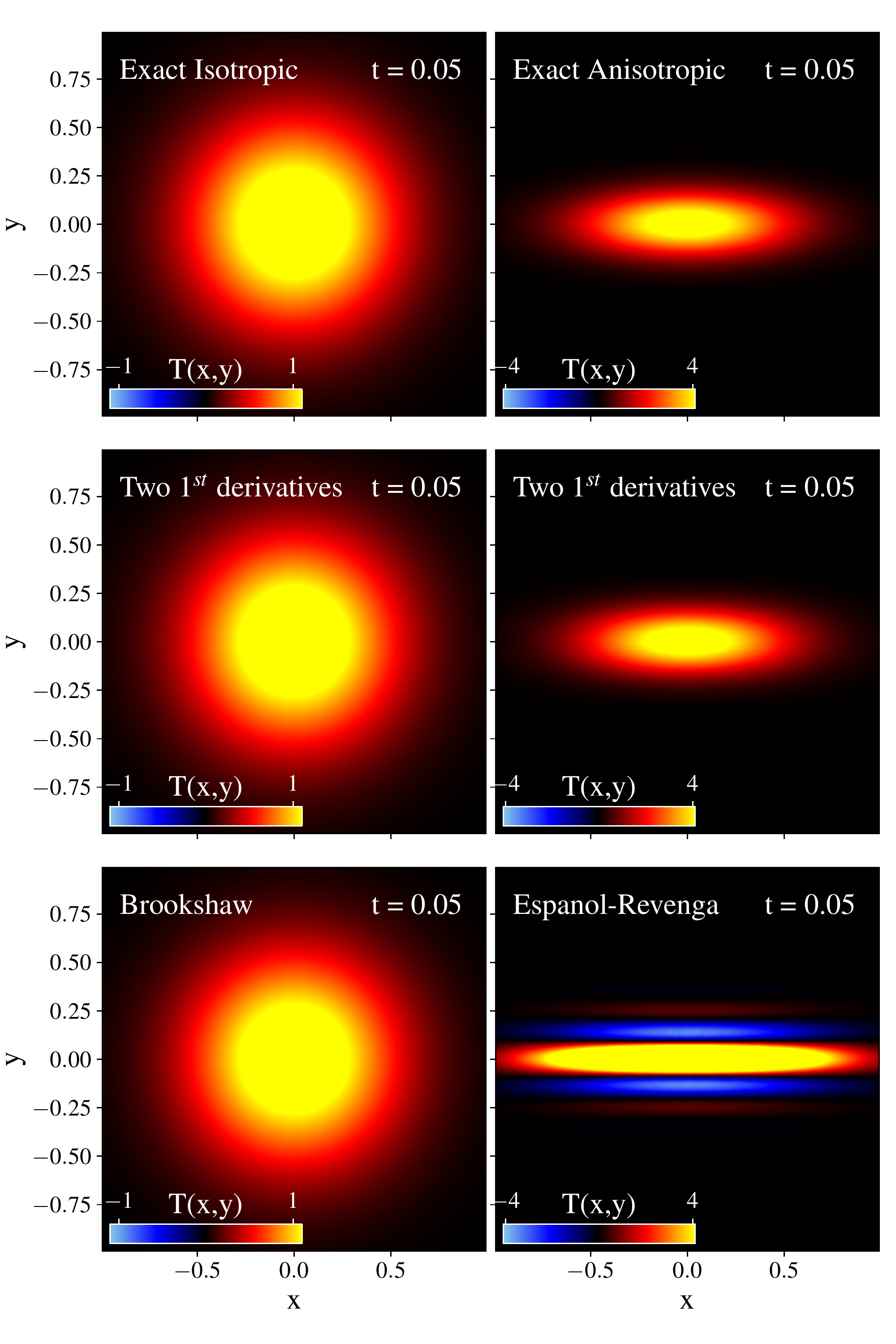}
    \captionof{figure}{Anisotropic diffusion in a 3D slab with constant heat conduction tensor. We use 64$^2$ particles placed uniformly and the $M_6$ kernel. At $t=0$ we assume a Gaussian pulse in the middle of the plate with $\epsilon = 0.1$. The only non-zero component of heat conduction tensor is $\kappa^{xx}$. Here we compare isotropic (left panels) and anisotropic (right panels) diffusion. In the isotropic case both methods agree with the exact solution with errors of order $\approx10^{-2}$. For the anisotropic heat conduction, the Espanol-Revenga method becomes unstable as described in Section~\ref{subsubsec_esanolrevengastability}, while two first derivatives retain the same convergence rate and accuracy (see Section~\ref{sec_convergence}).
    }
    \label{fig_gpk100}
  \end{center}
\end{figure}

\begin{figure}
  \begin{center}
    \includegraphics[width=1\linewidth]{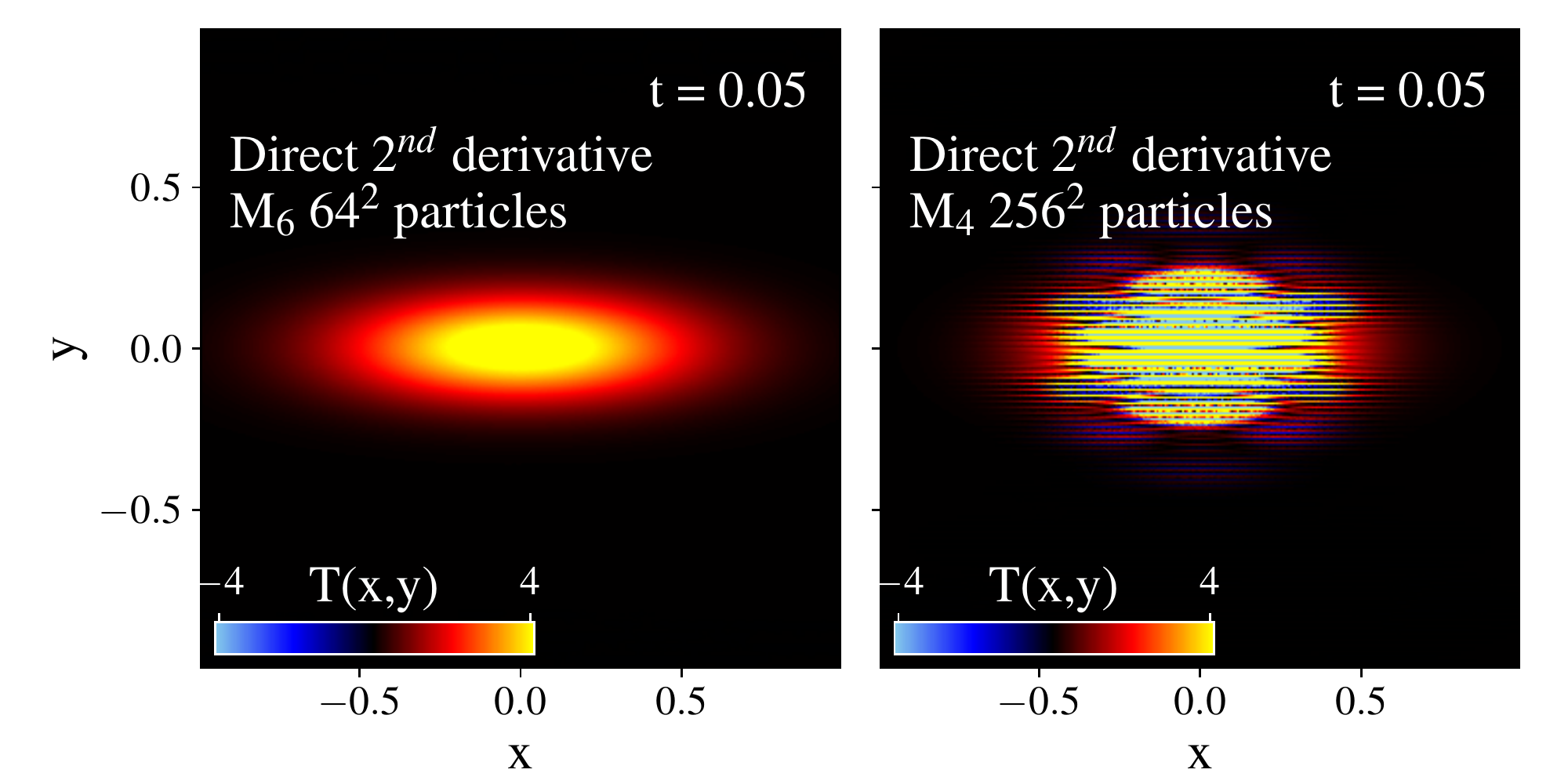}
    \captionof{figure}{As in Figure~\ref{fig_gpk100} but solved with the direct $2^{nd}$ derivative method. Although it is possible to obtain an accurate solution that appears stable (left panel), the instability reveals itself at higher resolution and for noisier kernels (right panel), as discussed in Section~\ref{subsubsec_directsecondderivsstability}. In this particular case, the problem becomes unstable after $t\approx0.025$ with the direct second derivative method when solved with the $M_4$ kernel on 256$^2$ particles.
    }
    \label{fig_gpan2w}
  \end{center}
\end{figure}

Figure~\ref{fig_gpk100} shows the results for isotropic diffusion using the Brookshaw method and for anisotropic diffusion using the Espanol-Revenga method (bottom row). The corresponding exact solutions are shown in the top row, respectively. The results highlight the instability inherent in the Espanol-Revenga method when the diffusion is highly anisotropic.
The solutions to the same problem performed with two first derivatives are stable (middle row).

It is possible to obtain accurate results at this resolution (64$^2$ particles) using the direct second derivative method as well, provided one employs the quintic kernel. However, the method does not guarantee stability and may become unstable, as demonstrated in Figure~\ref{fig_gpan2w}.

\subsection{3D diffusion with variable heat conduction tensor}
\label{subsec_difftest3dvarkappa}

If we consider the previous problem in cylindrical coordinates $(\rho, \phi, z)$, as shown by \cite{2017_Hopkins_anis_dif}, we can define the initial heat source according to
\begin{equation}
  T(\rho,\phi,0) = T_0 + T_1 \exp{\left[-\frac{1}{2}\left(\frac{(\rho - \rho_0)^2}{\delta \rho^2} + \frac{\phi^2}{\delta \phi^2} \right)\right]},
\end{equation}
where $\rho_0=0.3$ is a radial position of the source, $\delta \rho = 0.05$ is a radial size of the source, and $\delta \phi = 0.5$ is a azimuthal size of the source, $T_0=0$, $T_1=1$.

Then, the solution is given by
\begin{equation}
  T(\rho,\phi,t) = T_0 + T_1(t) \exp{\left[-\frac{1}{2}\left(\frac{(\rho - \rho_0)^2}{\delta \rho^2} + \frac{\phi^2}{\delta \phi_0^2 + 2 \kappa_{\phi\phi} \rho^{-2}t} \right)\right]}.
\end{equation}
If we set the heat conduction tensor to have only one nonzero component $\kappa_{\phi\phi}=1$, it means that there should be the only diffusion in the angular direction. The corresponding tensors in both cylindrical and cartesian coordinates are given by
\begin{equation}
  \boldsymbol{\kappa}_{\rho\phi z} =
  \begin{bmatrix}
    0 & 0 & 0 \\
    0 & 1 & 0 \\
    0 & 0 & 0
  \end{bmatrix}
; \boldsymbol{\kappa}_{xyz} = \frac{1}{x^2 + y^2}
  \begin{bmatrix}
    x^2 & -xy & 0 \\
    -yx & y^2 & 0 \\
    0 & 0 & 0
  \end{bmatrix}
\end{equation}
Another possible approach here would be to transform the formulae for SPH derivatives into cylindrical coordinates. We found similar results regardless of the method employed.

\begin{figure}
  \begin{center}
    \includegraphics[width=1\linewidth]{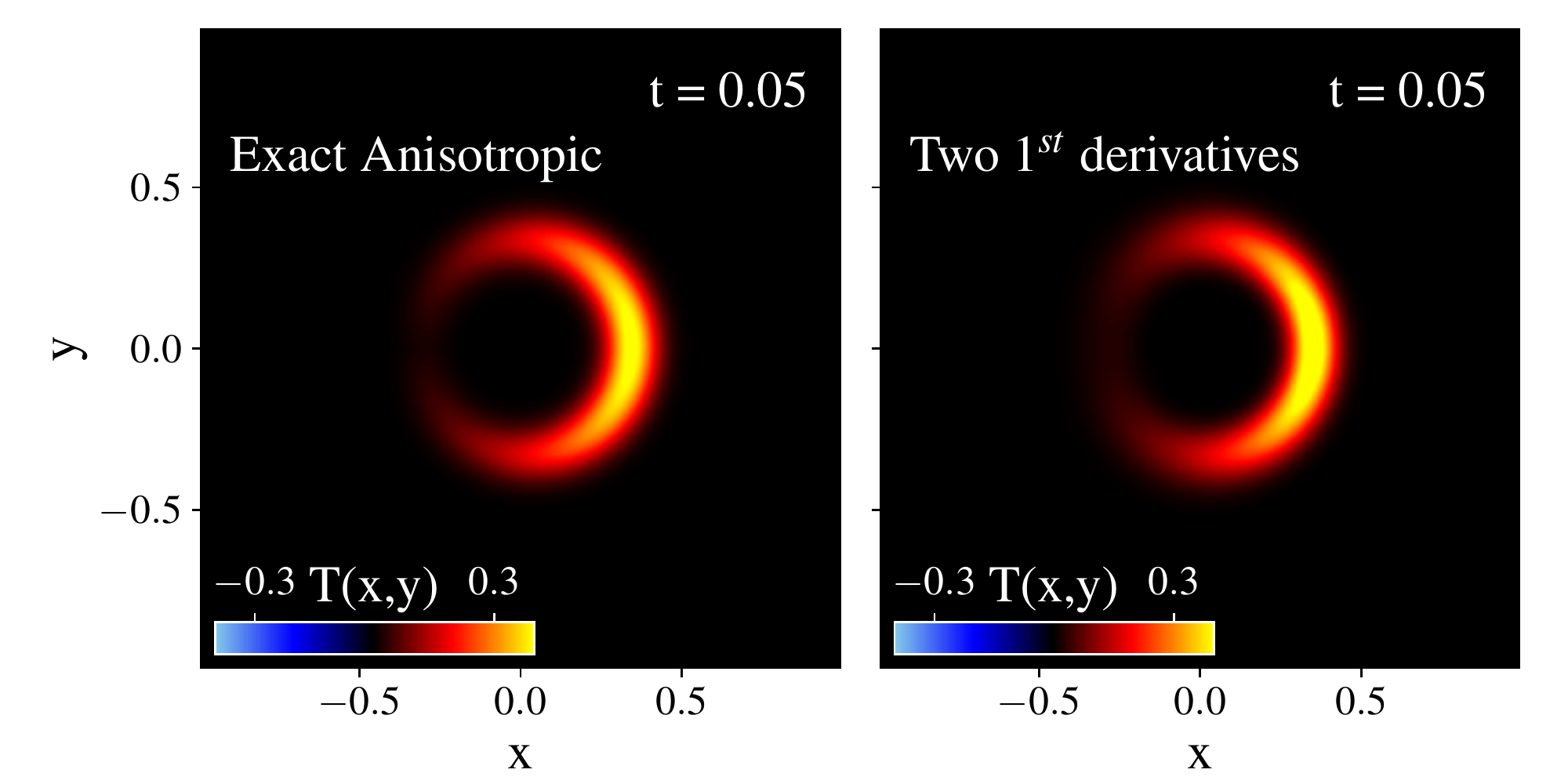}
    \captionof{figure}{
    \label{fig_grk100}
    Anisotropic diffusion in 3D slab with a spatially variable heat conduction tensor. At $t=0$ the heat is a Gaussian pulse with $\rho_0=0.3$, $\delta\rho=0.05$, and $\delta\phi=0.5$. The heat conduction tensor in cylindrical coordinates contains only one non-zero component $\kappa^{\phi\phi}=1$. Solution was obtained with 64$^2$ particles and the $M_6$ quintic kernel.}
  \end{center}
\end{figure}

Figure~\ref{fig_grk100} shows that two first derivative method can handle the situation where the direction of the heat flow is not aligned with the particle distribution. For the same problem, the Espanol-Revenga method results in numerical instability. The direct derivative method is an acceptable, but risky, choice, as it may become unstable depending on the particle arrangement, the number of neighbours, or kernel involved (we used the $M_6$ quintic spline here). By contrast, the proof of increasing entropy we gave in \ref{subsubsec_twofirststability} for the two first derivatives method does not depend on any of the above.

\subsection{Convergence and kernels}
\label{sec_convergence}

\begin{figure*}
  \begin{center}
    \includegraphics[width=1\textwidth]{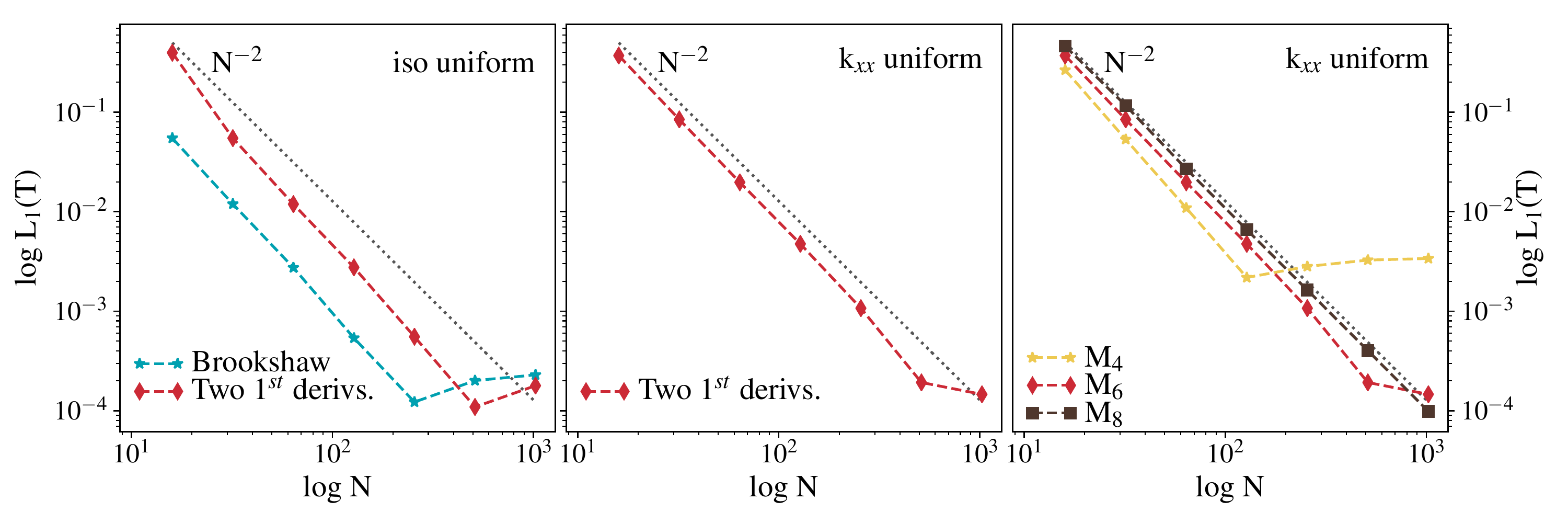}
    \caption{
    Convergence of error for isotropic and anisotropic diffusion of a Gaussian pulse (Section~\ref{subsec_difftest3dconstkappa}). Red dashed line shows the results for two first derivatives, while the blue dashed line corresponds to Brookshaw method. On the uniform lattice with continuous initial conditions, both methods converge quadratically (the black dotted line shows the expected slope) until the effect of kernel bias dominates at a resolution $\sim 10^3$ particles per direction. The kernel bias can be further eliminated by using smoother kernels with larger compact support radii (right panel).
    }
    \label{fig_conv_gp}
  \end{center}
\end{figure*}

\begin{figure*}
  \begin{center}
    \includegraphics[width=1\textwidth]{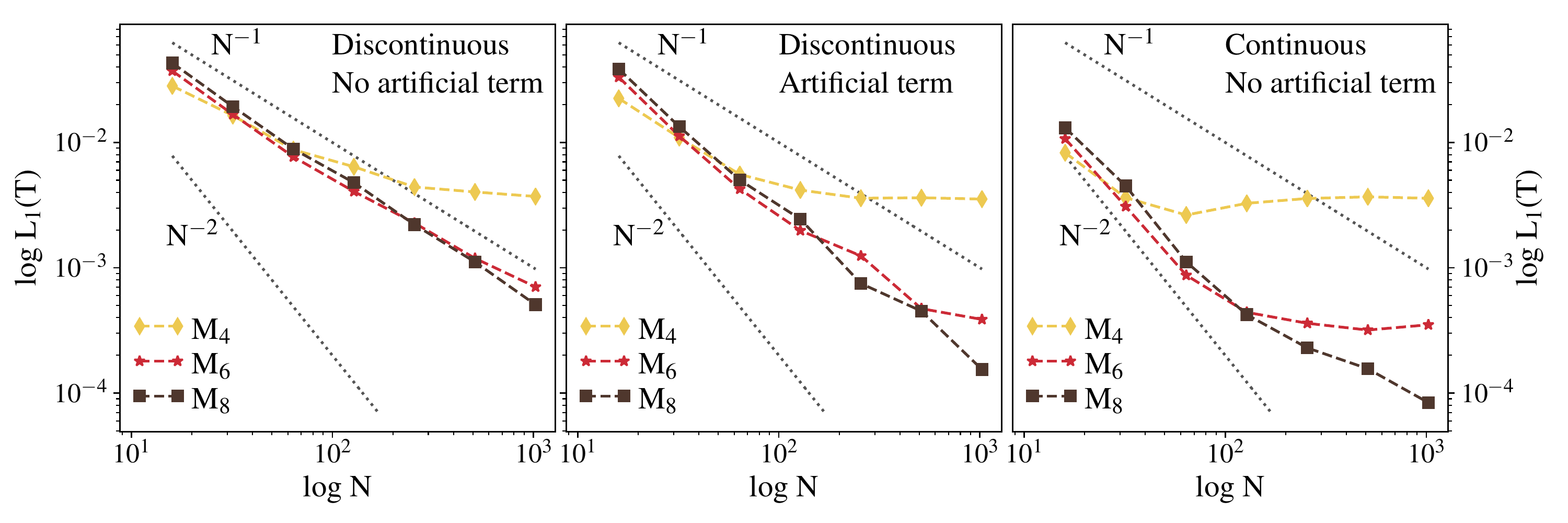}
    \caption{
    Convergence for anisotropic diffusion with a discontinuity in the initial conditions. We used the glass-like lattice and the two first derivatives method. Here we compare how the initial discontinuity influences the convergence properties. With no special treatment the convergence is linear (left panel). Adding artificial thermal conduction term improves this to $\sim 1.5$ (central panel). With continuous initial conditions, the error converges quadratically (right panel). The kernel bias remains the same across all methods for treating the discontinuity, and independent of the particle placement (convergence on a uniform lattice can be found on the right panel of Figure~\ref{fig_conv_gp}).
    }
    \label{fig_conv_hc12}
  \end{center}
\end{figure*}

 For each the preceding tests, we performed a convergence study, shown in Figures~\ref{fig_conv_gp} and \ref{fig_conv_hc12}. These show that the kernels and initial particle distribution have a significant influence on the overall accuracy. The kernel bias is the region in which the accuracy cannot be increased by simply using more particles.

 Figure~\ref{fig_conv_gp} (left panel) compares the convergence properties of the Brookshaw method and two first derivatives for isotropic heat conduction with particles on a regular lattice with the Gaussian pulse as the initial temperature distribution and quintic spline (M$_6$) as a smoothing kernel. Convergence is quadratic for both methods, even though the Brookshaw method is one order of magnitude more accurate overall. When the error reaches $\approx 10^{-4}$, it stops improving with higher resolution. The error then depends purely on the smoothing kernel. The central panel of the Figure~\ref{fig_conv_gp} shows that the convergence rate and kernel bias stay the same when we switch to the anisotropic problem (where the Brookshaw method is no longer applicable). The right panel of Figure~\ref{fig_conv_gp} shows that the choice of kernel and the number of neighbours sets the kernel bias. Using higher kernels in the spline series, we can obtain progressively more accurate results  at high spatial resolution.

 Figure~\ref{fig_conv_hc12} shows the convergence results for one dimensional heat diffusion at $t=0.025$ (Section~\ref{subsec_1ddifftest}). We adopted a glass-like lattice for different kernels and applied different methods to deal with the discontinuous initial conditions. Importantly, we see that the carbuncle mode is \emph{not} an instability, since the introduced noise decreases with resolution even if left without any treatment (left panel). The convergence is linear. When we use the artificial thermal conduction term (Eq.~\ref{eq_diffusion_art_term}) similar to the one used in hydrodynamics, the order of convergence increases to $\sim1.5$ (central panel). Finally, from the right panel, the convergence of the same problem with continuous initial conditions is second order once again.

Comparing Figures~\ref{fig_conv_gp}~and~\ref{fig_conv_hc12} demonstrates that the kernel biases remain similar independent of the initial conditions, particle placement, whether or not diffusion is anisotropic, and whether or not a discontinuity is present. For the $M_6$ spline kernel it is $\sim~5\times 10^{-4}$.

\section{Discussion}
\label{sec_discussion}

Our main finding is that the second law of thermodynamics is the most important consideration when assessing the stability of diffusion operators in SPH. This consideration leads to the conclusion that the only stable operator for anisotropic heat conduction in SPH is the `two first derivatives' method described in Section~\ref{subsubsec_twofirstderivs}. In the same time, we emphasised that it is crucial to alternate between differential and symmetric operators when constructing a two first derivatives method, in order to preserve positive increase of entropy. This conclusion was already reached by \cite{2017_Price_Phantom} in the context of physical viscosity and dust-gas mixtures. Similarly, the need for conjugate derivative operators to satisfy conservation laws is common in SPH \citep[e.g.][]{1992_monaghan_sph,1999_cummins_conjoper,2012_tricco_price}.

Likewise, the two first derivatives method has already been used widely, e.g. for dust-gas mixtures \cite{2015_price_laibe}, physical viscosity \citep{1994_Flebbe_PhysVisc,2006_springel_physvisc}, resistive and ambipolar diffusion \citep{2014_wurster_price_ayliffe} and the Hall effect \citep{2016_wurster_price_bate}. In the above papers, the authors found the method to be the most reliable.

For the case of anisotropic diffusion, our findings give an alternative to the fix suggested by \cite{2009_petkova_springel}. In particular, using two first derivatives does not require limiting the anisotropy of the flow in order to achieve stability. The \cite{2003_espanol_revenga} method should not be used for anisotropic diffusion.

The main caveat to using two first derivatives is the appearance of carbuncle modes if the initial conditions are discontinuous. The problem disappears with smooth initial conditions. Another solution is to introduce additional artificial conduction term that acts only when the solution (e.g. in pressure) is discontinuous, as proposed by \cite{2008_Price_discontSPH}. These terms are similar to the diffusion that would arise in a Finite Volume scheme when fluxes are reconstructed at discontinuous interfaces. 
The `integral Godunov' methods proposed by \cite{2017_Hopkins_anis_dif} indeed require flux limiters to ensure that the entropy is positive.
\cite{2017_Hopkins_anis_dif} pointed out that symmetric operators, in general, can result in low-order convergence if the particles are disordered.
We found this can be mitigated by the use of smoother spline kernels.
We found the \cite{2003_espanol_revenga} method to be significantly less accurate.

An interesting follow-up would be to search for a kernel function that satisfies condition~(\ref{eq_dirderinstabilitycondition}). If such a kernel exists, it may be possible to take direct second derivatives that are both stable and give non-oscillating solutions. An interesting application beyond the scope of this paper would be to the Magneto-Thermal Instability \citep[see][]{2017_Hopkins_anis_dif}.

\section{Conclusions}
\label{sec_conclusion}
We analysed the stability of methods for anisotropic diffusion in SPH. Our conclusions are:
\begin{enumerate}
  \item
  In case of isotropic diffusion the Brookshaw method is stable and also the most accurate method.
  \item Two first derivatives are the only method for anisotropic diffusion where stability is guaranteed.
  The only caveat is that smoothing or an artificial diffusion term is required for accurate treatment of discontinuities.
  \item We recommend against the use of the Espanol-Revenga method for anisotropic diffusion because it is not only unstable under certain circumstances but also inaccurate.
  \item
  We find that use of the M$_6$ quintic spline kernel reduces the kernel bias by approximately one order of magnitude compared to the cubic spline kernel for both isotropic and anisotropic diffusion.
  \item
  For the two first derivatives method one can use a timestep 3--8 times larger (depending on the choice of kernel) than for the Brookshaw method, while remaining stable. This offers a potentially large cost saving.
\end{enumerate}

\section*{Acknowledgements}
 We thank the organisers of SPHERIC2018 for a useful and stimulating conference in Galway, Ireland. DP acknowledges funding from the Australian Research Council via FT130100034 and DP180104235.

%
%
%
\bibliographystyle{mnras}
\bibliography{/Users/sergeibiriukov/_git/moca_study/latex/bibbase/BibDesk} 



%

\bsp    
\label{lastpage}
\end{document}